\documentclass[12pt,a4paper]{article}
\usepackage[cp866]{inputenc}
\usepackage{amsmath}
\usepackage{axodraw}
\usepackage[dvips]{epsfig}
\usepackage{graphicx}
\usepackage{epsfig}
\newcommand{\be}{\begin{equation}}
\newcommand{\ee}{\end{equation}}
\newcommand{\ben}{\begin{equation*}}
\newcommand{\een}{\end{equation*}}
\newcommand{\bea}{\begin{eqnarray}}
\newcommand{\eea}{\end{eqnarray}}
\newcommand{\ar}{\begin{array}}
\newcommand{\arn}{\end{array}}

\newcommand{\vk}{\vec{k}}
\newcommand{\vks}{\vec{k}^{\;2}}
\newcommand{\q}{\vec{q}}
\newcommand{\qs}{\vec{q}^{\;2}}
\newcommand{\qp}{\vec{q}^{\;\prime}}
\newcommand{\qps}{\vec{q}^{\;\prime\; 2}}

\newcommand{\vl}{\vec{l}}

 \def\eps{\epsilon}

\def\pnot{\mbox{${\not{\hbox{\kern-3.0pt$p$}}}$}}
\def\qnot{\mbox{${\not{\hbox{\kern-2.0pt$q$}}}$}}
\def\enot{\mbox{${\not{\hbox{\kern-2.0pt$e$}}}$}}
\def\knot{\mbox{${\not{\hbox{\kern-2.0pt$k$}}}$}}

\def\fun#1#2{\lower3.6pt\vbox{\baselineskip0pt\lineskip.9pt\ialign
{$\mathsurround=0pt#1\hfil##\hfil$\crcr#2\crcr\sim\crcr}}}

\begin{document}
%\numberwithin{equation}{section}     %\numerate of equation (first equation in each section)
\sloppy                              
\renewcommand{\baselinestretch}{1.0} %\constitute the line difference

\begin{titlepage}
\hskip 11cm \vbox{ \hbox{Budker INP 2016-23}  }
\vskip 3cm
%\vspace{1cm}\\

\begin{center}
{\bf Particularities of the NNLLA BFKL $^{\ast}$}
\end{center}

\centerline{V.S.~Fadin
$^{a,b\,\dag}$}

\vskip .6cm

\centerline{\sl $^{a}$
Budker Institute of Nuclear Physics of SD RAS, 630090 Novosibirsk
Russia}
\centerline{\sl\sl $^{b}$ Novosibirsk State University, 630090 Novosibirsk, Russia}

\vskip 2cm

\begin{abstract}
Peculiar properties of the BFKL  approach   in the next-to-next-to-leading logarithmic approximation (NNLLA) are discussed.   In this approximation the scheme of derivation of the BFKL equation must be changed because of  violation of the simple factorized form of amplitudes with  multi-Reggeon exchanges and necessity to take into account imaginary parts of amplitudes in the unitarity relations.
     \end{abstract}

%\vskip .5cm

\vfill \hrule \vskip.3cm \noindent $^{\ast}${\it Work supported %in part by INTAS,
in part by the Ministry of Education and Science of Russian Federation,
%grant  14.740.11.0082 of Federal Program "Personnel of Innovational Russia",
in part by  RFBR,  grant 16-02-00888.} 
\vfill $
%\begin{array}{ll} 
^{\dag}\mbox{{\it e-mail address:}}
% &
\;\;\; \mbox{fadin@inp.nsk.su}
%\\
%\end{array}
$

\end{titlepage}

\section{Introduction}
The BFKL (Balitsky-Fadin-Kuraev-Lipatov) approach~\cite{Fadin:1975cb,Kuraev:1976ge,Kuraev:1977fs,Balitsky:1978ic} provides a general way for the theoretical description of  processes at high  c.m.s. energy  $ \sqrt s $ and fixed (not growing with $s$) momentum transfers.  It  is based on a remarkable property of QCD -- gluon Reggeization~\cite{Lipatov:1976zz}, which gives a very powerful tool for the description of such processes.
This property makes  it possible to derive  the equation of evolution of the amplitudes of processes with energy (the BFKL equation). Now the BFKL approach is well developed in the leading  logarithmic approximation (LLA), which summarizes the terms $(\alpha_s \ln s)^n$ and in the next after it (NLLA),  giving the opportunity to summarize also the terms $\alpha_s (\alpha_s \ln s)^n$ (see e.g.~\cite{Ioffe:2010zz} and references therein). Development of the next (NNLLA) approximation is a long-standing problem, actual both in terms of phenomenology and theory. There is no doubt that such a development is possible. It turns out, however,  that the scheme of derivation of the BFKL equation must be changed in the NNLLA.

\section{The scheme of derivation of the BFKL equation}
The BFKL equation is derived by analyzing the $s$-channel discontinuities of elastic amplitudes calculated using unitarity. The main contributions to the discontinuities   are given by the multi-Regge kinematics (MRK).  Due to the Reggeization,  the amplitudes used in the unitarity conditions have a simple factorized form in the  NLLA  as well as in the LLA.    The  Reggeization  allows  to express an infinite number of amplitudes   through several effective vertices and gluon trajectory. For  elastic scattering processes  $A+B$ $\rightarrow
A^{\prime }+B^{\prime }$   the Reggeization  means that   scattering amplitudes with gluon quantum numbers in the $t$-channel in the Regge kinematic
region  $s \simeq - u \rightarrow \infty$, $t$ fixed,  can be presented as
\be
{{\cal A}}^{A' B'}_{AB} = {\Gamma}^R_{A^\prime
A} \left[\left({-s\over -t}\right)^{\omega(t)}-\left({s\over
-t}\right)^{\omega(t)}\right] \Gamma^R_{B^\prime B}~, \label{elastic}
\ee
where  ${\Gamma}^R_{P^\prime P}$ are energy-independent particle-particle-Reggeon (PPR) vertices (or  scattering vertices),  $j(t) =1 + \omega(t)$ is the  Reggeized gluon trajectory.
The Reggeization means also definite (multi-Regge) form   of  production  amplitudes in the multi-Regge kinematics  (MRK). MRK is the kinematics  where all particles have
limited (not growing with $s$) transverse momenta and are combined into jets with limited invariant mass of each jet and large (growing with $s$) invariant masses of any pair
of the jets.  This kinematics gives dominant contributions to cross sections of
QCD processes at high energy $\sqrt s$.

For the amplitude  ${\cal A}_{2\rightarrow n+2}$ of  the  process $A+B\rightarrow A'+G_1+\ldots+G_n+B'$  of production of $n$ gluons with momenta $k_1, k_2, \ldots k_n $
the  MRK means
\be
s \gg s_i \gg |t_i|\simeq \qs_i,~~~~ \;\; s\simeq \frac{ \prod_{i=1}^{n+1}s_i} {\prod_{i=1}^{n}\vks_i}~, \label{MRK}
\ee
where
\be
s=(p_A+p_B)^2, \; s_i=(k_{i-1}+k_i)^2, \;\;i=1, \cdot\cdot\cdot n+1, \;   \; k_{0}\equiv P_{A'}, \;\;k_{n+1}\equiv P_{B'},\label{notation1}
\ee
\be
q_1=p_A-p_A',\;   q_{j+1} = q_{j}-k_j, \; j=1, \cdot\cdot\cdot n, \; \; q_{n+1} =p_{B'} -p_B~,  \label{notation2}
\ee
the vector sign means transverse to the   $p_A, p_B$ plane  components.
In this region, one has for the amplitude  ${\cal A}_{2\rightarrow n+2}$
\be
\Re {\cal A}_{2\rightarrow n+2}=2s\Gamma^{{R}_1}_{
A' A} \left( \prod_{i=1}^{n}
 \frac{1}{t_{i}}\Big(\frac{s_i}{|\vk_{i-1}||\vk_{i}|}\Big)^{\omega(t_i)}
\gamma^{G_i}_{R_i{R}_{i+1}}\right)\frac{1}{t_{n+1}}\Big(\frac{s_{n+1}}{|k_{n}||\q_{n+1}|}\Big)^{\omega(t_{n+1})}
\Gamma^{{R}_{n+1}}
_{B' B}~.
\label{A 2-2+n}
\ee
Here ${\Gamma}^{{R}}_{ A'A}$ and $\Gamma^{{R}} _{B' B}$ are the same scattering vertices  as in (\ref{elastic})    and  $\gamma ^{G_{i}}_{{R}_i {R}_{i+1}}$ are the Reggeon-Reggeon-Particle (RRP) vertices (or the production vertices).

In the LLA only gluons  can be produced. In the NLA  one has to consider not only the  amplitudes~(\ref{elastic}),  (\ref{A 2-2+n}),   but also amplitudes obtained from them by  replacement of one of final particles by  a  jet containing a couple of particles with fixed (of order of transverse momenta) invariant  mass.

The Reggeon vertices and the gluon trajectory are known  in the next-to-leading order (NLO), that means the one-loop approximation for the vertices and the two-loop approximation for the trajectory. It is just the accuracy which is required for the derivation of the BFKL equation in the NLLA.
Validity of the forms  (\ref{elastic}) and (\ref{A 2-2+n} is proved now   in all orders of perturbation theory  in the coupling constant $g$ both  in the LLA~\cite{Balitskii:1979} and in the  NLLA~\cite{Fadin:2006bj,Kozlov:2011zza,Kozlov:2012zz,Kozlov:2012zza,Fadin:2015zea}. Note that the simple factorized  form~(\ref{A 2-2+n}  is valid only for the real part of the amplitudes (the sign $\Re $ in the Equation~(\ref{A 2-2+n}) means ``real part'').  Fortunately, the imaginary parts  are  not essential  for  the derivation of the BFKL equation in the NLLA.

The Reggeization provides a simple derivation of the BFKL equation both in the LLA and NLLA. Two-to-two scattering amplitudes  with all possible quantum
numbers in the $t$--channel are calculated using Equation~(\ref{elastic}) and (\ref{A 2-2+n}) in the  $s$-channel unitarity  relations  and analyticity.
The $s$-channel discontinuities  of the processes  $A+B\rightarrow A'+B'$ are presented as the convolutions $\Phi _{A^{\prime }A}\;\otimes \;G\;\otimes \;\Phi _{B^{\prime }B}$,
where the impact factors $\Phi _{A^{\prime}A}$ and  $\Phi _{B^{\prime }B}$ describe  transitions $A\rightarrow A^{\prime}$ and  $B\rightarrow B^{\prime }$ due to interactions  with Reggeized gluons, $G$ is the  Green's function for two
interacting Reggeized gluons with  an operator form ${\hat{\cal
G}}=e^{Y\hat{\cal{K}}}$, where $\;Y=\ln(s/s_0)$, $s_0$  is an energy scale, ${\hat{\cal{K}}}$ is the  BFKL kernel. The impact factors and the BFKL kernel   are expressed in terms of the Reggeon vertices and trajectory. Energy dependence of scattering amplitudes is determined by the  BFKL kernel, which is universal (process independent). The kernel ${\hat{\cal {K}}}={\hat{\cal
{\omega}_1}}+{\hat{\cal {\omega}}_2}+ {\hat{\cal{K}}_r}$
is expressed through the Regge trajectories ${\hat{\cal{\omega}}_1}$ and ${\hat{\cal {\omega}}_2}$ of two gluons  and the ``real part" ${\hat{\cal {K}}_r}$ describing production of particles in their interaction: ${\hat{\cal {K}}_r} ={\hat{\cal {K}}_G}+{\hat{\cal {K}}_{Q\bar Q}}+{\hat{\cal {K}}_{GG}}$. In the LLA only  ${\hat{\cal {K}}_G}$ must be kept, because only  gluons can be produced; in the NNLLA production of quark-antiquark (${Q\bar Q}$)  and gluon ($GG$) pairs  is  also possible.

One might think that this scheme is applicable in the NNLLA as well. In this case it would be sufficient to calculate three-loop corrections to the trajectory, two-loop corrections to ${\hat{\cal {K}}_G}$, one-loop corrections to ${\hat{\cal {K}}_{Q\bar Q}}$ and ${\hat{\cal {K}}_{GG}}$ and  to find in the Born approximation two new contributions,  ${\hat{\cal {K}}_{Q\bar Q G}}$ and ${\hat{\cal {K}}_{GGQ}}$,  to ${\hat{\cal {K}}_r}$.

Unfortunately, the scheme based on the forms~(\ref{elastic})  and~(\ref{A 2-2+n})
does not work in the NNLLA. The reason is the need  to take account of the contributions of Regge cuts and the imaginary parts of the amplitudes in the  unitarity conditions.

\section{Contributions of the three-Reggeon cut}
In the NLLA, two large logarithms can be lost in the product of two amplitudes in the unitarity condition used for derivation of the BFKL equation. It can be done  losing either both logarithms in one of the amplitudes, or  one logarithm in each of the amplitudes. In the first case  one of the amplitudes is taken in the NNLLA and the other in the LLA. Since the amplitudes in the LLA are real, only real parts of the NNLLA amplitudes are important in this case. But even for these parts the forms~(\ref{elastic}) and~(\ref{A 2-2+n}) become inapplicable because of the contributions of the three-Reggeon cut  that appear in this approximation.

The first observation of the violation of the form (\ref{elastic})  was made~\cite{DelDuca:2008pj}  in the consideration of the high-energy limit of the two-loop amplitudes for quark-quark ($qq$), quark-gluon ($qg$) and gluon-gluon ($gg$) scattering. The interference of the tree- and two-loop amplitudes for each of the  processes has been explicitly computed.  The discrepancy appears in non-logarithmic two-loop  terms. If the form~(\ref{elastic}) would be correct in the NNLLA, they should satisfy  the certain relation, because in (\ref{elastic}) three amplitudes are expressed in terms of only two vertices $\Gamma^R_{QQ}$ and  $\Gamma^R_{GG}$.
The explicit calculation  gives that  this relation is violated by terms of ${\cal O}(\pi^2/\eps^2)$.

Detailed consideration of  the terms  responsible  for the violation of the factorized form (\ref{elastic})  in the case of two-loop and three-loop quark and gluon amplitudes was performed in~\cite{DelDuca:2013ara,DelDuca:2013dsa,DelDuca:2014cya}.  In particular, the  non-logarithmic double-pole contribution at two-loops  obtained in~\cite{DelDuca:2008pj}  was recovered  and all non-factorizing single-logarithmic singular contributions at  three loops were found using the  techniques of infrared factorization.

All these results are explained by the three-Regge cut contributions~\cite{FL:2016}.  Since our Reggeons are the Reggeized gluons, the cut starts from the diagrams with three-gluon exchanges. In the Feynman gauge the  contribution of these diagrams to the amplitudes $A^{a}$  with the adjoint representation of the colour group in the $t$-channel has the form
 \be
A_{ij}^a=\langle A'|T^a|A\rangle \langle B'|T^a|B\rangle\left[ C_{ij} A^{(eik)} + \frac{N_c^2}{8} \left(A_{ij}^{s}+A_{ij}^{u}\right)+ \delta_{i,q}\delta_{ij,q}\frac{4-N_c^2}{8} \left(A_{ij}^{s}-A_{ij}^{u}\right)\right]~, \label{Aij}
\ee
where  $ij$ are $qq, qg$ and $gg$ for quark-quark, quark-gluon and gluon-gluon scattering correspondingly, $A_{ij}^{s}$ and $A_{ij}^{u}$ are the the contributions of the ladder diagrams in the $s$ and $u$ channels respectively with omitted colour group factors,  $A^{(eik)}$ is the sum  of such  contributions  for all the diagrams, and $C_{ij}$ are the colour group coefficients:
\be  
C_{qq} =\frac14\left(-1+\frac{3}{N_c^2}\right)~,\;\; C_{qg} =\frac14~,\;\; C_{gg} =\frac32~. \label{Cij}
\ee 
The last term in~(\ref{Aij})  is the contribution of the positive signature in the quark-quark scattering and is imaginary.  The second term has the form~(\ref{elastic}) and can be assigned to the Reggeized gluon contribution.  On the contrary, this is not true for  the first term, because
\be
2C_{gq} \neq  C_{qq}+C_{gg}~.
\ee
Exactly this term is responsible for the violation of the factorized form~(\ref{elastic})  at  the  two-loop level   discovered in~\cite{DelDuca:2008pj} and confirmed in~\cite{DelDuca:2013dsa}. 
This follows from the explicit form  $A^{eik} = g^2({s}/{t})({-4\pi^2}/{3})g^4\,\qs\,A^{(3)}_\perp $, 
\be
A^{(3)}_\perp \ =\  \int\frac{{d^{2+2\epsilon}l_1}{d^{2+2\epsilon} l_2}}{(2\pi)^{2(3+2\epsilon)}\vec l_1^{\:
2}\vec l_2^{\:
2}(\vec q -\vec l_1-\vec l_2)^2}
= 3C_{\Gamma}^2\frac{4}{\epsilon^2}\frac{(\vec q^{\:
2})^{2\epsilon}}{\vec q^{\:
2}} \frac{\Gamma^2(1+2\epsilon)\Gamma(1-2\epsilon)}
{\Gamma(1+\epsilon)\Gamma^2(1-\epsilon)\Gamma(1+3\epsilon)}~,
\ee
\be
C_{\Gamma}= \frac{\Gamma(1-\epsilon)\Gamma^2(1+\epsilon)}
{(4\pi)^{2+\epsilon}\Gamma(1+2\epsilon)}~.\;\;\;
\ee
However, one can not affirm that this term  is given entire by the three-Reggeon cut. Indeed, the  coefficients $C_{ij}$ in  (\ref{Cij}) can be presented  as the sum
\be
C_{ij} =C^R_{ij} + C^C_{ij}, \label{Cij as sum}
\ee
with the coefficients $C^R_{ij}$ satisfying the equality 
\be
2C^R_{qg} =   C^R_{qq}+C^R_{gg}~. 
\ee
The terms with $C^R_{ij}$ have the form~(\ref{elastic}) and can be assigned to the Reggeized gluon contribution, so that  the contribution of the three-Reggeon cut can be given by the terms with $C^C_{ij}$ only. Since the coefficients $C^R_{ij}$ obey only one condition, there is a great freedom in their choice. It occurs~\cite{FL:2016} that
\be
C^R_{gg} = 3~, \;  C^C_{gg} = -\frac32~, \;C^R_{gq} =\frac{7}{4}~, \; \;C^C_{gq} =-\frac{3}{2}~, \;  C^R_{qq} =\frac{1}{2}~, \; \;C^C_{qq} =\frac{3(1-N_c^2)}{4N_c^2}~. 
\label{coefficients CRij and CCij}
\ee
The Reggeon and three-Reggeon cut contributions have different dependence on $s$. In the case of the Reggeized gluon it come  solely from the Regge factor as in~(\ref{elastic}).  In the case of the three-Reggeon cut, one has to take into account the Reggeization of each  of  the three gluons and the interaction between them. For the first logarithmic correction,  the Reggeization gives $\ln s$ with the coefficient  $3C_R$, where
\be
C_R =  -g^2 N_c C_{\Gamma}\frac{4}{3\epsilon}  (\vec q^{\:
2})^{\epsilon}
\frac{\Gamma(1-3\epsilon)\Gamma(1+2\epsilon)\Gamma(1+3\epsilon)}
{\Gamma(1-\epsilon)\Gamma(1-2\epsilon)\Gamma(1+\epsilon)\Gamma(1+4\epsilon)}~. \label{CR}
\ee
Interaction between two  Reggeons with transverse momenta $\vl_1$ and $\vl_2$  and colour indices $c_1$ and $c_1$is given by the real part of the BFKL kernel
\be
\left[{\cal K}_r(\q_1, \q_{2}; \vk)\right]^{c'_1c'_2}_{c_1c_2} =
T_{c_1c'_1}^{a}T_{c_2c'_2}^{a}\frac{g^2}{(2\pi)^{D-1}}
\left[\frac{\qs_{1}\qps_{2}+\qs_{2}\qps_{1}}
{\vks}-\qs\right], \label{explicit K r}
\ee
where  $\vk$ is the  momentum transferred  from one Reggeon to another in the  interaction, $\qp_1$  and $\qp_2$ ($c'_1$ and $c'_2$) are the Reggeon momenta (colour indices) after the interaction, $\qp_1=\q_1 -\vk, \;\; \qp_2=\q_2 +\vk, $  and $\q =\q_1+\q_2=\qp_1+\qp_2$.

It occurs that, for the colour structure which we are interested in,  account of interactions between all  pairs  of the Reggeons   leads to the sum of  the colour coefficients which  differ from the coefficients $C_{ij}$ (\ref{Cij})
only by the common factor $N_c$. Therefore, the first order correction in the case  of the three-Reggeon cut is presented as  $(-4C_R -C_3 )\ln s$, where
the  $C_R$ and $C_3$ come correspondingly from the first two terms and the last term in the square brackets in  (\ref{explicit K r}),
\be
C_3 =    g^2 N_c C_{\Gamma}\frac{32}{9\epsilon}  (\vec q^{\:
2})^{\epsilon}
\frac{\Gamma(1-3\epsilon)\Gamma(1-\epsilon)\Gamma^2(1+3\epsilon)}
{\Gamma^2(1-2\epsilon)\Gamma(1+2\epsilon)\Gamma(1+4\epsilon)}~.\label{C3}
\ee
Thus, the first order correction  in the case of the three-Reggeon  cut  is $(-C_R -C_3)\ln s$, where $C_R$ and $C_3$ are given by (\ref{CR}) and  (\ref{C3}) respectively, and in the case of Reggeized gluon is $\omega(t)\ln s$, where
\be
\omega(t)
 =-g^2 N_c {\vec q^{\:
  2}} \int\frac{d^{2+2\epsilon}l}{2(2\pi)^{(3+2\epsilon)}\vec l^{\:
2}(\vec q -\vec l)^2}= -g^2 N_c C_{\Gamma}\frac{2}{\epsilon}(\vec q^{\:
2})^{\epsilon} ~.
\ee
With the colour coefficients~(\ref{Cij as sum}),  (\ref{coefficients CRij and CCij}), the terms singular in $\epsilon$ of the  total correction agree with the result obtained in~\cite{DelDuca:2014cya}.

Evidently, the three-Reggeon cut gives contributions to all $2\to n+2$ amplitudes in the MRK. They must be also found for further development of the   BFKL approach.

\section{Account of imaginary parts}
The sign $\Re $ in the Equation~(\ref{A 2-2+n}) means the ``real part''. It is important that the simple factorized  form (\ref{A 2-2+n})  is valid only for the real part of the amplitudes. Fortunately, the imaginary parts  are  not essential  for  the derivation of the BFKL equation in the NLLA. Indeed,  they are  suppressed by one power of $\ln s_i$ in comparison with the real ones,  and products of imaginary and real parts in the unitarity relations cancel  due to summation of contributions complex conjugated to each other.   But their account  becomes  necessary  in the NNLLA.

Fortunately, the  imaginary parts are needed only in the main approximation, so that their calculation  is not associated with large computational difficulties. However, it complicates the derivation of the BFKL equation and  deprives its universality,
because consideration of quark-quark, quark-gluon and gluon-gluon scattering becomes different.  It is clear already from consideration of two-particle intermediate states in the  unitarity condition.  Remind that for two Reggeized gluons in the $t$-channel in QCD, that is for three colours, there are 6 irreducible representations:
$\underline 1, \underline{8_a}, \underline{8_s},
\underline{10}, \underline{\overline{10}}, \underline{27}$ (for  $N_c > 3$ there is  one additional representation). The representations $\underline{8_a},
\underline{10}, \underline{\overline{10}}$ are anti-symmetric, while
the representations   $ \underline 1,  \underline{8_s}, \underline{27}$ (and the extra one for  $N_c > 3$) are symmetric. In real parts, with the NLLA accuracy, only the Reggeon channel,  $\underline{8_a}$, is  important. It provides universality of the NLLA: $gg, qg$ and $qq$ scattering can be considered  in an unique way.

But account imaginary parts violate the universality, because
$gg$ scattering amplitudes can contain all the representations, while
$qg$ and $qq$ amplitudes only $\underline 1, \underline{8_a}, \underline{8_s}$.

Consideration of many-particle states in the unitarity condition
is an even more complicated problem.

\section{Summary}
The basis of the BFKL approach is the remarkable property of
QCD --- gluon Reggeization.  In this approach amplitudes of elastic scattering are restored  analytically  from the imaginary parts calculated using unitarity. The main contributions to the imaginary parts  in the unitarity conditions come from the multi-Regge kinematics (MRK).  In the leading (LLA) and next-to-leading (NLLA) logarithmic approximations  the Reggeization provides with required accuracy  a simple factorized form of QCD amplitudes used in the unitarity conditions.   In the NNLLA  such form is violated by the three-Reggeon cut   and by imaginary parts of the amplitudes. For further  development of the BFKL approach the Regge cut contributions  must be found. Contributions of the  three--Reggeon cut  to  elastic amplitudes were found in~\cite{FL:2016} and are presented here. Account of this cut in inelastic amplitudes is under consideration.  As regards  the  imaginary parts required  in the unitarity conditions, the way of their calculation  is known.  Unfortunately, the need to take them into account violates universality of the derivation of the BFKL equation for various processes, as well as the need to incorporate the Regge cuts.

\section{Acknowledgments}
I thank the Dipartimento di Fisica
dell'Universit\`{a} della Calabria and the Istituto Nazionale di
Fisica Nucleare (INFN), Gruppo Collegato di Cosenza, for warm hospitality
while part of this work was done and for financial support.


\begin{thebibliography}{99}
%1==================BFKL approach
\bibitem{Fadin:1975cb}%BFKL
V.S.~Fadin, E.A.~Kuraev, and L.N.~Lipatov,
%``On The Pomeranchuk Singularity In Asymptotically Free Theories'',
Phys.\ Lett.\  B {\bf 60},  50-52 (1975).
%2================== BFKL approach
\bibitem{Kuraev:1976ge}%BFKL
E.A.~Kuraev, L.N.~Lipatov, and V.S.~Fadin,
%``Multi - Reggeon Processes In The Yang-Mills Theory,''
  Zh. Eksp. Teor. Fiz. \textbf{71},   840-855 (1976) [Sov. Phys. JETP \textbf{44},
443-450 (1976)].
%3================== BFKL approach
\bibitem{Kuraev:1977fs}%BFKL
E.A.~Kuraev, L.N.~Lipatov, and V.S.~Fadin,
%``The Pomeranchuk Singularity In Nonabelian Gauge Theories,''
Zh.\ Eksp.\ Teor.\ Fiz.\ {\bf 72},   377-389  (1977) [Sov.\ Phys.\ JETP {\bf 45},  199-207  (1977)].
%4================== BFKL approach
\bibitem{Balitsky:1978ic}
  I.I.~Balitsky,  and L.N.~Lipatov,
  %``The Pomeranchuk Singularity In Quantum Chromodynamics,''
 Yad.\ Fiz.\  {\bf 28},   1597-1611 (1978)
  [Sov.\ J.\ Nucl.\ Phys.\  {\bf 28},   822-829 (1978)].
%5-----------------------------------------------------------------------
\bibitem{Lipatov:1976zz}
  L.N.~Lipatov,
  % ``Reggeization Of The Vector Meson And The Vacuum Singularity In Nonabelian
  %Gauge Theories,''
Yad.\ Fiz.\  {\bf 23}, 642-656 (1976)
[Sov.\ J.\ Nucl.\ Phys.\  {\bf 23},  338-345 (1976)].
%6-----------------------------------------------------------------------
\bibitem{Ioffe:2010zz}
  B.~L.~Ioffe, V.~S.~Fadin and L.~N.~Lipatov,
  %``Quantum chromodynamics: Perturbative and nonperturbative aspects,''
  Cambridge University Press,  2010,   ISBN: 9781107424753
%7-----------------------------------------------------------------------
\bibitem{Balitskii:1979}
Ya.Ya.~Balitskii, L.N.~Lipatov, and V.S.~Fadin,{\it ``Regge Processes In Nonabelian Gauge Theories"} (in Russian), in {\it Materials of
IV Winter School of LNPI} (Leningrad, 1979), pp. 109-149.
%8------------------------------------------------------------------------
\bibitem{Fadin:2006bj}
V.S.~Fadin, R.~Fiore, M.G.~Kozlov, and A.V.~Reznichenko,
%``Proof of the multi-Regge form of QCD amplitudes with gluon exchanges in
%the NLA,''
Phys.\ Lett.\  B {\bf 639},   74-81  (2006). %[hep-ph/0602006].
%9------------------------------------------------------------------------
%\bibitem{Fadin:2002et}
% V.~S.~Fadin,
  %``Justification of the BFKL approach in the NLA,''
%  "Diffraction 2002", Ed. by R. Fiore {\it et al}., NATO Science Series, Vol. 101, pp.235-245.
%10-----------------------------------------------------------------------
\bibitem{Kozlov:2011zza}
  M.G.~Kozlov, A.V.~Reznichenko, and V.S.~Fadin,
  %``Check of the gluon-Reggeization condition in the next-to-leading order: Quark part,''
 Yad.\ Fiz.\  {\bf 74}, 784-796  (2011)
   [Phys.\ Atom.\ Nucl.\  {\bf 74},  758-770 (2011)].
%11-----------------------------------------------------------------------
\bibitem{Kozlov:2012zz}
 M.G.~Kozlov, A.V.~Reznichenko, and V.S.~Fadin,
  %``Impact factor for gluon production in multi-Regge kinematics in the next-to-leading order,''
Yad.\ Fiz.\  {\bf 75}, 905-920   (2012)
  [Phys.\ Atom.\ Nucl.\  {\bf 75},   850-863 (2012)].
%11-------------------------------
\bibitem{Kozlov:2012zza}
  M.G.~Kozlov, A.V.~Reznichenko, and V.S.~Fadin,
  %`` Check of the gluon-Reggeization condition in the next-to-leading order: Gluon part,''
 Yad.\ Fiz.\  {\bf 75}, 529-542   (2012) [Phys.\ Atom.\ Nucl.\  {\bf 75},  493-506 (2012).
%11-----------------------------------
%\bibitem{Kozlov:2013zza}
%  M.G.~Kozlov, A.V.~Reznichenko, and V.S.~Fadin,
  %``Multi-Regge form of  amplitudes with  gluon exchanges  in  supersymmetric Yang-Mills theories,
%Yad.\ Fiz.\  {\bf 77},  273-294 (2014)  [Phys. At. Nucl. {\bf 77}, 251-273 (2014)].
%12-----------------------------------
\bibitem{Fadin:2015zea}
  V.~S.~Fadin, M.~G.~Kozlov,  and A.~V.~Reznichenko,
  %``Gluon Reggeization in Yang-Mills Theories,''
  Phys.\ Rev.\ D {\bf 92}  no.8,  085044 (2015).
%13-----------------------------------
\bibitem{DelDuca:2008pj}
  V.~Del Duca and E.~W.~N.~Glover,
  %``Testing high-energy factorization beyond the next-to-leading-logarithmic accuracy,''
  JHEP {\bf 0805}, 056 (2008)
%14-----------------------------------
\bibitem{DelDuca:2013ara}
  V.~Del Duca, G.~Falcioni, L.~Magnea and L.~Vernazza,
  %``High-energy QCD amplitudes at two loops and beyond,''
  Phys.\ Lett.\ B {\bf 732}, 233 (2014).
%15-----------------------------------
\bibitem{DelDuca:2013dsa}
  V.~Del Duca, G.~Falcioni, L.~Magnea and L.~Vernazza,
  %``Beyond Reggeization for two- and three-loop QCD amplitudes,''
  PoS RADCOR {\bf 2013}, 046 (2013).
% [arXiv:1312.5098 [hep-ph]].
  %%CITATION = ARXIV:1312.5098;%%
%16-----------------------------------
\bibitem{DelDuca:2014cya}
  V.~Del Duca, G.~Falcioni, L.~Magnea and L.~Vernazza,
  %``Analyzing high-energy factorization beyond next-to-leading logarithmic accuracy,''
  JHEP {\bf 1502}, 029 (2015).
%17-----------------------------------
\bibitem{FL:2016}
  V.S. Fadin, and L.N. Lipatov, in preparation.
%==================
\end{thebibliography}
\end{document}